%% file: kite.tex
\documentclass{article}
\usepackage{arxiv}
\usepackage{times}
\usepackage{natbib}
\setcitestyle{authoryear,round,citesep={;},aysep={,},yysep={;}}
\usepackage[T1]{fontenc}

\input{math_commands.tex}

\usepackage{hyperref}
\usepackage{url}
\usepackage{enumitem}
\usepackage{graphicx}
\usepackage{booktabs}
\usepackage{multirow}
\usepackage{siunitx}
\usepackage{tabularx}

\usepackage{listings}
\usepackage{tcolorbox}
\tcbuselibrary{listings,breakable,skins}

\definecolor{kitelink}{RGB}{28,62,132}
\hypersetup{colorlinks=true,linkcolor=kitelink,citecolor=kitelink,urlcolor=kitelink}

\lstdefinestyle{kitepromptstyle}{
    language={},
    basicstyle=\normalfont\ttfamily\normalsize,
    columns=fullflexible,
    keepspaces=true,
    breaklines=true,
    breakatwhitespace=true,
    breakautoindent=true,
    breakindent=1em,
    showstringspaces=false,
    showspaces=false,
    showtabs=false,
    upquote=true,
    numbers=none,
    tabsize=2,
    aboveskip=0pt,
    belowskip=0pt
}

\newtcblisting{kiteprompt}[1]{
    enhanced jigsaw,
    breakable,
    listing only,
    listing engine=listings,
    listing options={style=kitepromptstyle},
    title={#1},
    title after break={#1\ (continued)},
    fonttitle=\normalfont\bfseries\normalsize,
    colback=white,
    colframe=black!55,
    colbacktitle=black!4,
    coltitle=black,
    boxrule=0.4pt,
    sharp corners,
    boxsep=0pt,
    left=6pt,
    right=6pt,
    top=5pt,
    bottom=5pt,
    toptitle=4pt,
    bottomtitle=4pt,
    before skip=6pt,
    after skip=8pt,
    lines before break=2,
    width=\linewidth
}

\title{KITE: Retrieval as a Decision Procedure for Long-Term Conversational Memory}

\author{
  Feng Li$^{1}$ \quad Kailun Tan$^{1}$ \quad Zhibo Zhou$^{2}$ \quad Youxiang Wang$^{1}$ \\
  \textbf{Lejun Liu$^{1}$ \quad Huaigu Li$^{1,3}$ \quad Fei Du$^{4}$ \quad Huilong An$^{1,*}$} \\[6pt]
  {\normalfont $^{1}$Memoket AI Research \quad $^{2}$Universit\"{a}t Stuttgart \quad $^{3}$Georgia Institute of Technology \quad $^{4}$Tsinghua University} \\
  {\normalfont $^{*}$Corresponding author: \texttt{dev@memoket.ai}}
}
\date{}

\renewcommand{\headeright}{Preprint}
\renewcommand{\undertitle}{Preprint}
\renewcommand{\shorttitle}{KITE: Retrieval as a Decision Procedure for Long-Term Conversational Memory}

\begin{document}

\maketitle

\begin{abstract}

Many long-term memory systems for LLM agents are built around embedding-based retrieval.
Although effective for semantic matching, similarity ranking alone does not establish whether retrieved memories satisfy the conditions implied by a question.
In this paper, we introduce \textsc{KITE} (\textbf{K}nowledge-\textbf{I}ndexed \textbf{T}emporal \textbf{E}vidence), a vector-free framework that formulates memory retrieval as an explicit decision procedure.
\textsc{KITE} organizes multi-session conversations as structured knowledge indexed by a governed topic vocabulary, which evolves through controlled expansion and consolidation.
At query time, questions are compiled into inspectable retrieval plans over the same knowledge index that guides memory construction.
The plans are then executed through fixed lookup and ranking operations, while sparse lexical retrieval recovers surface details omitted from the structured representation.
Experiments on LoCoMo and LongMemEval-S show that \textsc{KITE} achieves 93.5\% and 86.0\% accuracy while using only about 1.6k reader-context tokens per question.
These results demonstrate that conversational memory can move beyond similarity ranking by unifying memory construction and query compilation around a governed index, turning retrieval into an inspectable decision procedure for determining which evidence satisfies the question.

\end{abstract}

\section{Introduction}
\label{sec:introduction}

As large language model (LLM) agents engage in multi-session interactions, they need to carry useful knowledge forward from earlier exchanges.
Although larger context windows expose more history, they do not inherently support selective retention \citep{hsieh2024ruler}.
Persistent memory has therefore become a key component of long-term LLM-agent architectures, allowing relevant information to be retained and reused across sessions \citep{zhong2024memorybank}.

\begin{figure}[h]
    \centering
    \includegraphics[width=\linewidth]{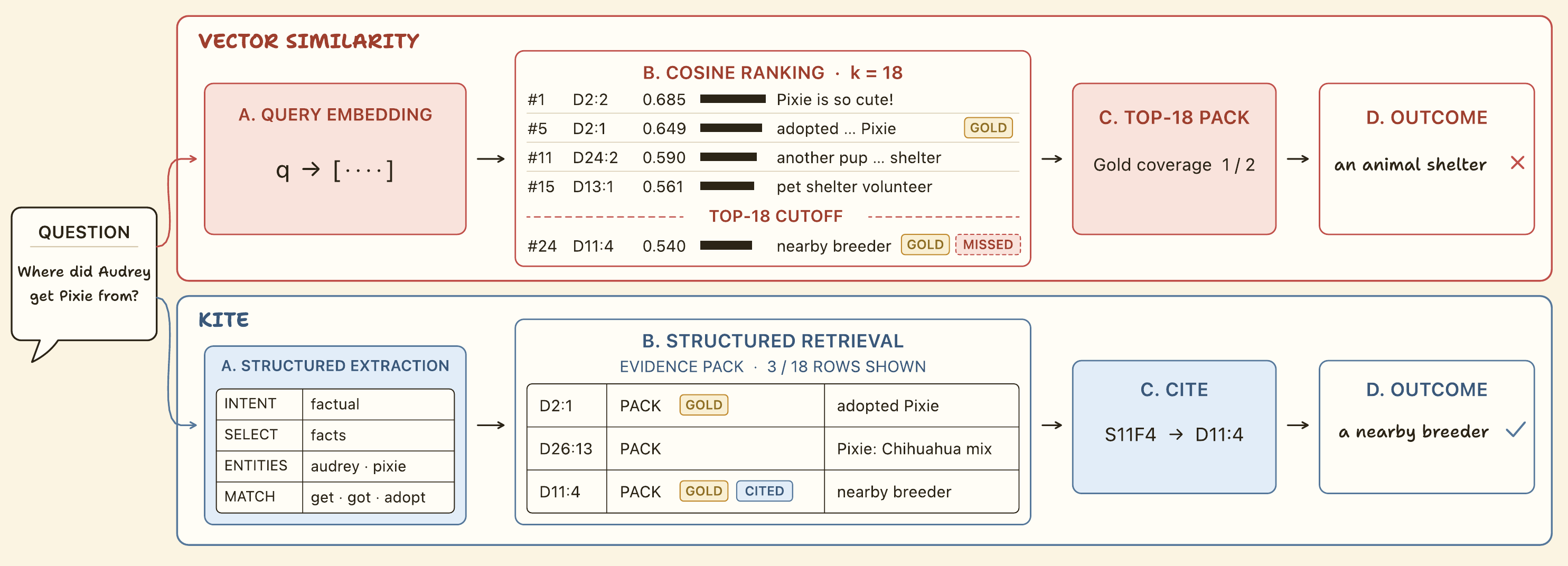}
    \caption{Comparison of embedding retrieval and \textsc{KITE}'s retrieval with explicit constraints.}
    \label{fig:retrieval-decision}
\end{figure}

In practice, many memory systems retrieve past information through nearest-neighbor search over embeddings \citep{packer2023memgpt, chhikara2025mem0}.
Although effective for semantic matching, single-vector similarity ranking alone does not ensure that the retrieved memories satisfy the conditions of the query \citep{weller2026theoretical}.
As Figure~\ref{fig:retrieval-decision} illustrates, semantically close memories may outrank evidence that answers the question \citep{trappolini2026redefining}.
In addition, fixed top-$k$ selection returns candidates without explicitly indicating when none satisfies the query \citep{kim2025speak}.
Beyond retrieval, storing $N$ memories as $d$-dimensional dense vectors requires $O(Nd)$ space, while encoder changes often require rebuilding the entire embedding index \citep{monath2024fresh, goswami2025query}.

These limitations of embedding-based retrieval motivate a formulation of memory retrieval as an explicit decision procedure that evaluates query conditions rather than relying on similarity ranking alone \citep{wang2026text2mem}.
By exposing matching conditions and intermediate results, this approach makes retrieval decisions inspectable and supports the diagnosis of retrieval failures \citep{ru2024ragchecker}.
However, as conversations evolve, the same concepts may be represented by different surface labels or graph nodes, complicating consistent indexing \citep{zhang2024extract, mo2026kggen}.
Moreover, questions may refer to stored information through different wording or implicit cues, making relevant evidence harder to retrieve \citep{hu2026querylink}.
These challenges call for a governed knowledge index that evolves with the ongoing conversation, together with query compilation that is grounded in the same knowledge representation.

To this end, we propose \textsc{KITE}\footnote{\url{https://github.com/memoket/memoket-kite}} (\textbf{K}nowledge-\textbf{I}ndexed \textbf{T}emporal \textbf{E}vidence), a vector-free framework for long-horizon conversational memory that organizes multi-session interactions into structured knowledge indexed by a governed topic vocabulary.
Each memory entry is enriched with typed contextual and temporal facets, providing complementary access paths beyond topics.
At query time, \textsc{KITE} compiles a natural-language question into an inspectable retrieval plan over the same governed knowledge index, specifying the conditions to apply.
The validated plan is then executed through explicit lookup and ranking operations, making evidence selection deterministic.
\textsc{KITE} also supplements the primary path with sparse lexical retrieval to recover surface details missed during memory construction.
Experiments on LoCoMo \citep{maharana2024evaluating} and LongMemEval-S \citep{wu2024longmemeval} show that \textsc{KITE} achieves accuracies of 93.5\% and 86.0\%, respectively, while remaining token-efficient by limiting reader context to a compact evidence set.

The key contributions of this paper are as follows:

\begin{itemize}[leftmargin=*, labelindent=0pt]
    \item \textbf{Retrieval as a deterministic decision procedure.}
    We formulate memory retrieval as a decision procedure that evaluates explicit query conditions against indexed evidence.
    Validated plans are then executed through fixed lookup and ranking operations, making evidence selection deterministic per plan and inspectable.
    These mechanisms establish a practical alternative to similarity-based retrieval while maintaining high accuracy with compact reader contexts.

    \item \textbf{A governed and evolving knowledge index.}
    We organize memories using a topic vocabulary that evolves across multi-session conversations under controlled expansion and consolidation, with typed facets for finer-grained access to evidence.
    This design limits topic fragmentation and keeps memories coherently organized as conversations grow.

    \item \textbf{A shared index for memory construction and query compilation.}
    We align question interpretation with memory organization via a shared knowledge index.
    Mapping question concepts to existing index keys grounds retrieval plans in stored topics and entities, reducing the semantic gap between natural language and indexed knowledge.
\end{itemize}

\section{Related Work}

\subsection{Long-Term Memory for LLM Agents}

Extending the context window of language models is a direct approach to supporting long-horizon interaction \citep{beltagy2020longformer, bulatov2022recurrent}.
However, longer contexts increase inference costs, while model performance remains susceptible to position-dependent degradation \citep{liu2024lost, hutter2025lost}.
Persistent-memory systems instead preserve information across sessions, beyond the limits of the active context window.
For example, MemGPT \citep{packer2023memgpt} manages information transfer between the model context and the external storage.
In addition, MemoryBank \citep{zhong2024memorybank} incorporates a time-dependent forgetting mechanism, whereas Mem0 \citep{chhikara2025mem0} extracts salient facts into an editable memory store and supports add, update, and delete operations.

Beyond individual memory operations, several works investigate broader mechanisms for managing information accumulated over long interactions.
A prominent direction explores system-level memory architectures with hierarchical storage and lifecycle control \citep{kang2025memory, li2025memos}.
Meanwhile, MEM1 \citep{zhou2026mem1} consolidates interaction history into a compact internal state that supports continued reasoning.
Together, they show that long-term memory depends not only on persistence, but also on how accumulated information is organized for later use.

\subsection{Memory Organization and Index Governance}

Recent studies have explored structured representations for organizing conversational memory.
A-MEM \citep{xu2026mem} builds interconnected notes whose attributes and links evolve as new memories arrive.
Similarly, Zep \citep{rasmussen2025zep} organizes conversational knowledge as a temporal knowledge graph that tracks facts along with their validity over time.
Alongside these relational approaches, Memanto \citep{abtahi2026memanto} structures memories through predefined semantic types, while EverMemOS \citep{hu2026evermemos} consolidates episodic MemCells into thematic MemScenes.
In general, these approaches provide richer structures for organizing related memories over time.

As these structures grow, a related challenge is maintaining consistent representations of recurring concepts.
EDC \citep{zhang2024extract} canonicalizes extracted relations into a consolidated schema, including settings without a predefined target schema.
KGGen \citep{mo2026kggen} addresses the same problem from a graph-construction perspective by clustering entities and relations to reduce fragmentation.
These works highlight the importance of keeping related concepts aligned and reducing redundant representations, providing a more stable basis for subsequent retrieval.

\subsection{Query Alignment and Executable Retrieval}

Once memories are organized, retrieval must connect each question to the representations in which evidence is stored.
Sparse lexical methods such as BM25 \citep{robertson2009bm25} rank candidates by scoring matches between query terms and stored text.
However, questions may express the same information in different words, making relevant memories harder to find through lexical matching alone.
QueryLink \citep{hu2026querylink} addresses this mismatch by building semantic representations of questions and memories at different levels of detail for similarity-based retrieval.

Complementary work makes memory access explicit through executable operations.
ChatDB \citep{hu2023chatdb} treats databases as symbolic memory, allowing an LLM to generate SQL instructions for memory retrieval and manipulation.
More generally, Text2Mem \citep{wang2026text2mem} translates natural-language memory commands into validated operations with a common execution interface.
For more structured knowledge, KAG \citep{liang2025kag} combines logical-form planning with hybrid retrieval over knowledge graphs and text, while supporting schema-free knowledge extraction.
Executable programs can further support reasoning after retrieval, as in TReMu \citep{ge2025tremu}, which performs temporal calculations over retrieved timeline summaries.
Taken together, these approaches establish executable operations as a general-purpose mechanism for explicit access and reasoning over stored evidence.

This emphasis on explicit operations also aligns with diagnostic work on intermediate evidence.
RAGChecker \citep{ru2024ragchecker} separates retrieval from generation to localize pipeline failures.
In long-term memory, MemTrace \citep{long2026memtrace} distinguishes whether source evidence is reachable from whether it is successfully used.
These evaluations highlight the value of making the evidence process inspectable rather than relying on final-answer accuracy alone.

\begin{figure}[t]
    \centering
    \includegraphics[width=\linewidth]{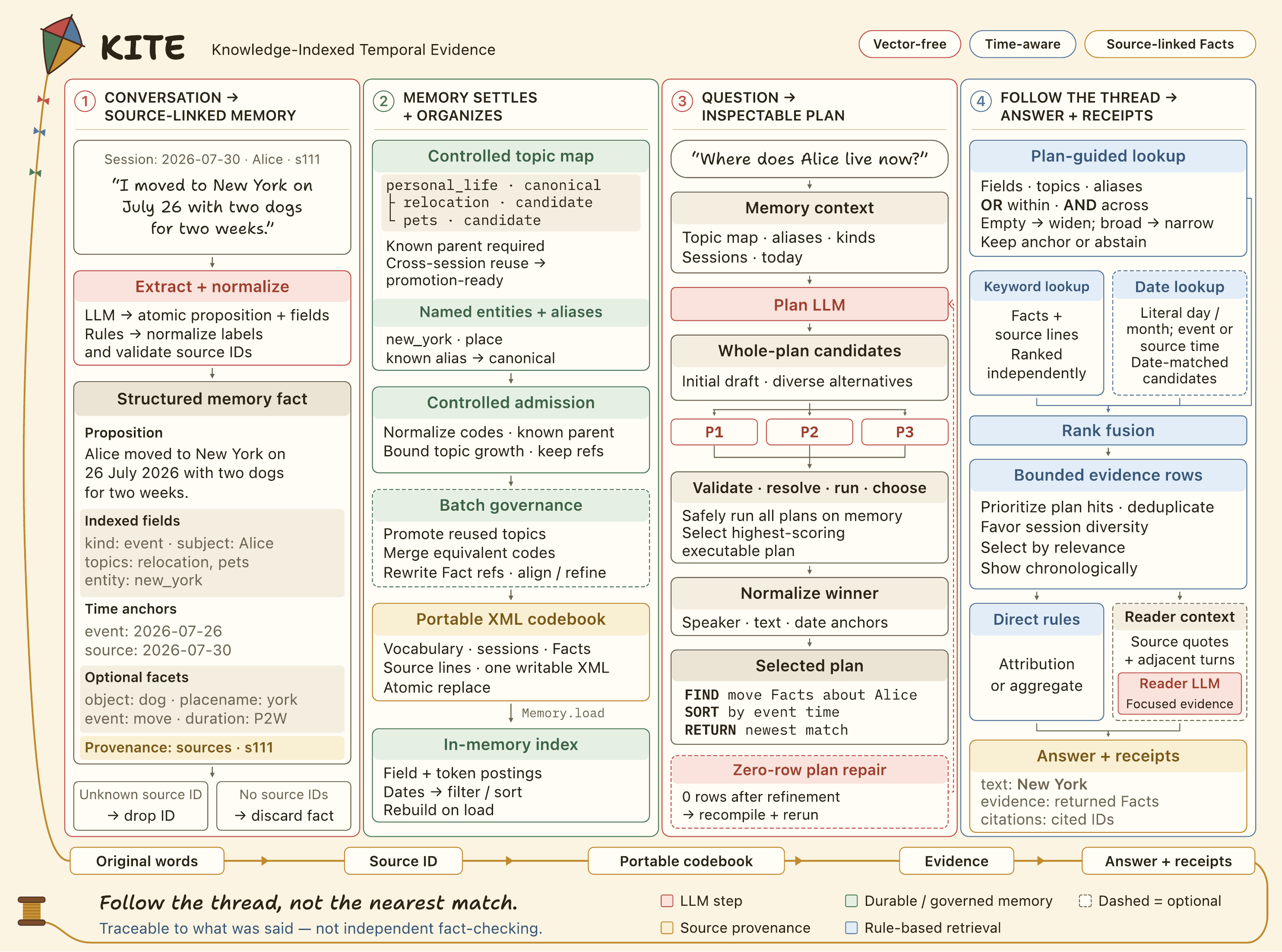}
    \caption{
        Overview of \textsc{KITE}.
        Multi-session conversations are converted into structured memory entries indexed by governed topics.
        Given a question, \textsc{KITE} compiles an inspectable retrieval plan over the same knowledge index to locate relevant contextual and temporal evidence for answering.
    }
    \label{fig:kite-overview}
\end{figure}

\section{Method}
\label{sec:method}

To connect memory organization with evidence selection, \textsc{KITE} formulates long-term conversational memory retrieval as an explicit decision procedure over a governed knowledge index.
As illustrated in Figure~\ref{fig:kite-overview}, each session is converted into structured memory entries indexed under a governed topic vocabulary and enriched with contextual and temporal facets.
At query time, \textsc{KITE} uses the same knowledge index to compile each question into an inspectable retrieval plan that specifies the conditions for selecting evidence.
The validated plan is then executed deterministically through fixed lookup and ranking operations, while sparse lexical retrieval provides complementary evidence.

Consider a sequence of dated conversation sessions $U=(u_1,\ldots,u_m)$ ordered chronologically.
Given a later natural-language question $q$, the retrieval task is to recover evidence from this history that supports an answer.
\textsc{KITE} constructs a set of structured memory entries $M$ from $U$ and selects relevant entries via a knowledge-indexed retrieval path.

Let $\Sigma$ be the set of governed index keys.
Memory indexing and query compilation are defined as
\begin{equation}
\mathcal{I}: M \rightarrow 2^{\Sigma},
\qquad
\mathcal{Q}_{\mathbf{P}}: \mathcal{X} \rightarrow \Pi ,
\label{eq:index-and-compile}
\end{equation}
where $\mathcal{X}$ is the space of natural-language questions and $\Pi$ is the space of retrieval plans.
For each $m\in M$, $\mathcal{I}(m)\subseteq\Sigma$ specifies its index keys, and these assignments form the knowledge index $\mathbf{P}$.
Given this index, $\mathcal{Q}_{\mathbf{P}}(q)=\pi$ compiles a query $q$ into a retrieval plan $\pi$ whose indexed references resolve to the same keys used to organize the stored memories.

After validation, the indexed path executes the plan $\pi$ and returns a ranked evidence sequence $E_q$ as
\begin{equation}
E_q
=
\mathcal{D}_{\mathbf{P},\theta}(\pi)
=
\operatorname{Rank}_{\theta}
\left(
\operatorname{Exec}_{\mathbf{P}}(\pi)
\right),
\label{eq:decision-procedure}
\end{equation}
where $\operatorname{Exec}_{\mathbf{P}}$ applies the plan over the knowledge index and $\operatorname{Rank}_{\theta}$ orders the resulting candidates according to a fixed policy $\theta$.
For fixed $\mathbf{P}$, $\pi$, and $\theta$, the ranked evidence sequence $E_q$ is deterministic, and the decision procedure returns no match instead of forcing unrelated candidates when no indexed evidence satisfies the validated plan.

\subsection{Governed Knowledge Index}
\label{sec:knowledge-index}

\paragraph{Structured memory entries and provenance.}
Rather than indexing raw dialogue directly, \textsc{KITE} converts each conversation session into structured memory entries.
Each entry can be represented as
\begin{equation}
m=\langle x_m,\rho_m,T_m,A_m\rangle,
\label{eq:memory-representation}
\end{equation}
where $x_m$ is the self-contained memory content and $\rho_m$ its provenance, including the source session and supporting utterances.
The set $T_m$ holds the entry's governed topics, while $A_m$ stores typed contextual and temporal facets, such as entities and events.
These fields separate evidence from the structured information used to index and retrieve it.

For each session, the extraction model processes the dialogue with the current topic vocabulary as context, producing source-linked entries with facet annotations.
Relative time expressions are also resolved against the date of the corresponding session whenever possible, while candidate memory entries without valid provenance are discarded.

\paragraph{Governed topic vocabulary.}
Rather than assuming a complete ontology in advance, \textsc{KITE} derives its topic vocabulary from the accumulated conversation history.
Starting from a small set of canonical roots, the extraction model reuses existing topics when possible and gradually proposes more specific children when new concepts emerge.
To ensure consistency, new topics must fit within the existing hierarchy, which helps maintain a stable index over time.

Vocabulary expansion is further controlled by keeping proposed topics provisional until they recur across multiple sessions and time periods, rather than admitting each new topic as soon as it first appears.
As the vocabulary evolves, potentially redundant sibling topics are consolidated, with their memory entries redirected to the selected canonical topic.
These controls reduce unnecessary topic fragmentation while preserving stable index keys for retrieval.

\paragraph{Faceted knowledge indexing.}
Topics provide broad thematic access to memory, while typed facets offer finer-grained conditions for locating relevant evidence.
The index space is defined as
\begin{equation}
\Sigma =
\Sigma_{\mathrm{topic}}
\cup \Sigma_{\mathrm{entity}}
\cup \Sigma_{\mathrm{participant}}
\cup \Sigma_{\mathrm{object}}
\cup \Sigma_{\mathrm{place}}
\cup \Sigma_{\mathrm{event}} .
\label{eq:index-key-space}
\end{equation}
A memory entry may therefore be assigned keys from multiple types, providing complementary access paths to the same evidence.
For example, consider the LoCoMo question \emph{In which state is the shelter from which James adopted the puppy?} \citep{maharana2024evaluating}.
The dialogue describes James adopting a puppy named Ned from a shelter in Stamford.
\textsc{KITE} indexes this entry with \texttt{James} as a participant, \texttt{Ned} as an entity, \texttt{dog} as an object, \texttt{Stamford} as a place, and \texttt{adoption} as an event.
Although the question does not explicitly mention \texttt{Ned}, the participant and event keys can provide access to the relevant entry, where the required location information can be recovered.

For each key type $a$, \textsc{KITE} constructs a posting list:
\begin{equation}
P_a[s]
=
\left\{
m\in M
\mid
s\in\mathcal{I}_a(m)
\right\},
\label{eq:posting-list}
\end{equation}
where $s\in\Sigma_a$ denotes an index key and $\mathcal{I}_a(m)$ gives the keys of type $a$ assigned to memory entry $m$.
Each posting list gives direct access to the memories under a key.
Temporal annotations are stored separately in chronological order, enabling time constraints to be evaluated during retrieval.
These structures form the governed knowledge index $\mathbf{P}$.

\subsection{Index-Grounded Query Compilation}
\label{sec:query-compilation}

\paragraph{Shared-index grounding.}
Query compilation uses the same knowledge index built during memory construction, so each question is interpreted against the stored vocabulary and entities.
Thus, references inferred from the question can be resolved to existing index keys, keeping query interpretation aligned with memory organization.
The compiler then maps the question to these keys, while the executor determines which memories satisfy the resulting conditions.

\paragraph{Typed retrieval plans.}
Each question is compiled into a typed retrieval plan that makes its evidence requirements explicit before execution.
Formally, the plan consists of one or more subqueries
\begin{equation}
\pi = (\pi_1,\ldots,\pi_n), \qquad
\pi_i =
\left\langle
\mathtt{select},
\mathtt{where},
\mathtt{pipe}
\right\rangle,
\quad i=1,\ldots,n .
\label{eq:typed-plan}
\end{equation}
Here, \texttt{select} determines the retrieval granularity for each subquery, while \texttt{where} specifies the indexed and temporal conditions on evidence.
Finally, \texttt{pipe} applies fixed operations to the results, such as sorting, limiting, or counting.

Subqueries can be linked sequentially, allowing results from an earlier step to constrain the search performed in subsequent stages.
These dependencies keep multi-hop retrieval explicit in the plan while preserving fixed execution.

\paragraph{Plan validation.}
Before execution, each plan is validated against a fixed typed grammar and the knowledge index.
Indexed references are resolved to valid keys in the governed representation.
Speaker mentions are mapped to participant constraints, while temporal expressions are normalized into explicit ranges.
Plans with unresolved references or unsupported structures are rejected.
The validated plan is executed by the retrieval executor without further model intervention, drawing a clear boundary between model-based interpretation and deterministic evidence selection.

\subsection{Retrieval as a Decision Procedure}
\label{sec:decision-retrieval}

\paragraph{Deterministic plan execution.}
Given a validated subquery $\pi_i$, \textsc{KITE} evaluates its indexed conditions through the corresponding posting lists.
Requested keys within the same type are treated as alternatives, whereas conditions across types must hold jointly.
The initial candidate set is therefore
\begin{equation}
C(\pi_i)
=
\bigcap_{a\in\mathcal{A}(\pi_i)}
\left(
\bigcup_{s\in\pi_{i,a}} P_a[s]
\right),
\label{eq:candidate-set}
\end{equation}
where $\mathcal{A}(\pi_i)$ denotes the set of constrained key types and $\pi_{i,a}\subseteq\Sigma_a$ contains the requested index keys of type $a$.
Temporal and lexical conditions further filter this initial candidate set before the \texttt{pipe} operations are applied.

In practice, the executor narrows down the set of candidate memories according to the conditions encoded in the retrieval plan.
For dependent subqueries, results from preceding steps are carried forward as constraints on later retrieval.
With these operations fixed, the same index and validated plan always yield the same candidate set.

\paragraph{Controlled relaxation and no-match.}
When exact execution produces no candidates, \textsc{KITE} applies a predefined sequence of relaxations.
Each step broadens only designated conditions while preserving bindings from prior retrieval stages.
The effective conditions remain explicit throughout, and execution stops at the first non-empty result.
If all permitted relaxations fail, the indexed path explicitly returns no match under the current index and execution policy.

\paragraph{Explicit evidence ranking.}
After condition evaluation determines the candidate set, \textsc{KITE} orders the retained memories using a fixed scoring policy
\begin{equation}
S_{\theta}(m,\pi)
=
\sum_{j\in\mathcal{J}}
w_j\,\phi_j(m,\pi),
\label{eq:explicit-ranking}
\end{equation}
where each $\phi_j(m,\pi)$ captures an interpretable match between memory entry $m$ and retrieval plan $\pi$, and $w_j$ is its corresponding weight under the scoring policy $\theta$.
In practice, exact matches on index keys receive the strongest preference, while broader hierarchical matches or agreement on additional typed facets provide supporting signals.

For example, a memory matching both the queried entity and event outscores one linked only through a broader topic.
Ranking is applied only after condition checks admit the candidates, so it changes their ordering rather than eligibility.
For fixed $\mathbf{P}$, $\pi$, and $\theta$, the scores are deterministic, while the basis of each ranking decision remains inspectable.

\paragraph{Sparse lexical support.}
Structured memory is designed to capture reusable concepts and retrieval cues, but this abstraction may omit surface-form details preserved in the source text, such as rare names or numerical values.
To retain access to such information, \textsc{KITE} augments the indexed path with a deterministic sparse lexical channel.
The ranked evidence from both paths is merged under a fixed fusion rule, without dense embeddings or neural reranking.

A retrieval attempt may fail to produce a structured match without failing to find anything at all.
In that case the result is recorded as a fallback, and the lexical channel may still provide evidence to the reader, marked as such rather than as a match.
The reader thus gains additional evidence while the meaning of a match remains unchanged.

\section{Experiments}
\label{sec:experiments}

\subsection{Experimental Setup}
\label{sec:experimental-setup}

\paragraph{Benchmarks.}
We evaluate \textsc{KITE} on two long-term conversational memory benchmarks, LoCoMo \citep{maharana2024evaluating} and LongMemEval-S \citep{wu2024longmemeval}.
Specifically, the LoCoMo experiments cover 1,540 questions across ten multi-session conversations, while LongMemEval-S comprises 500 questions over longer interaction histories.

\paragraph{Baselines.}
We compare \textsc{KITE} against prior memory systems, namely Mem0 \citep{chhikara2025mem0}, Zep \citep{rasmussen2025zep}, MemOS \citep{li2025memos}, and EverMemOS \citep{hu2026evermemos}.
We also consider Naive RAG as a retrieval baseline and full-context inference over the complete conversation history as a non-retrieval reference.

\paragraph{Evaluation protocol.}
We evaluate answer quality using LLM-as-a-judge accuracy, reported overall and by question category.
Throughout the \textsc{KITE} experiments, \texttt{gpt-4.1-mini} serves as both reader and judge.
The reader receives only the question and retrieved evidence, without the full conversation history, and answers strictly from this context.

\paragraph{Implementation details.}
Unless otherwise specified, \textsc{KITE} combines plan-guided retrieval over the knowledge index with a sparse lexical channel under a fixed fusion strategy.
We use small benchmark-specific seed vocabularies and apply the same governance procedure across both benchmarks.
Further implementation details are provided in Appendices~\ref{app:experimental-details}--\ref{app:prompts}, and the source code is available at \url{https://github.com/memoket/memoket-kite}.

\begin{table}[t]
\centering
\small
\setlength{\tabcolsep}{4.0pt}
\renewcommand{\arraystretch}{1.05}
\caption{Comparison of memory systems on LoCoMo using LLM-as-a-judge accuracy (\%).}
\label{tab:locomo}

\begin{tabular}{@{}lcccccc@{}}
\toprule
\multirow{2}{*}[-0.7ex]{\textbf{System}} & \multirow{2}{*}[-0.7ex]{\textbf{Avg.\ Tokens}} & \multicolumn{5}{c}{\textbf{Accuracy (\%)}} \\
\cmidrule(lr){3-7}
& & \textbf{Single} & \textbf{Multi} & \textbf{Temp} & \textbf{Open} & \textbf{Overall} \\
\midrule
Naive RAG & \textbf{0.7k} & 42.6 & 42.6 & 67.9 & 71.9 & 49.7 \\
Full Context & 20.3k & 94.5 & 90.1 & 88.2 & 70.8 & 90.9 \\
\addlinespace[3pt]
Mem0 & 1.0k & 69.0 & 61.7 & 58.3 & 50.0 & 64.2 \\
Zep & 1.4k & 90.8 & 81.9 & 77.3 & 75.0 & 85.4 \\
MemOS & 2.5k & 85.4 & 79.4 & 75.1 & 64.6 & 80.8 \\
EverMemOS & 2.3k & \textbf{96.7} & 91.8 & 89.7 & 76.0 & 93.1 \\
\midrule
\textsc{KITE} & 1.5k & 95.2 & \textbf{93.3} & \textbf{92.2} & \textbf{83.3} & \textbf{93.5} \\
\bottomrule
\end{tabular}

\vspace{2pt}
\parbox{\linewidth}{
\footnotesize
Avg.\ Tokens denotes average reader-context tokens. Best results are \textbf{bolded}.
}
\end{table}

\begin{table}[t]
\centering
\small
\setlength{\tabcolsep}{4.0pt}
\renewcommand{\arraystretch}{1.05}
\caption{Comparison of memory systems on LongMemEval-S using LLM-as-a-judge accuracy (\%).}
\label{tab:longmemeval}

\begin{tabular}{@{}lcccccccc@{}}
\toprule
\multirow{2}{*}[-0.7ex]{\textbf{System}} & \multirow{2}{*}[-0.7ex]{\textbf{Avg.\ Tokens}} & \multicolumn{7}{c}{\textbf{Accuracy (\%)}} \\
\cmidrule(lr){3-9}
& & \textbf{SS-U} & \textbf{SS-A} & \textbf{SS-P} & \textbf{Multi} & \textbf{Know} & \textbf{Temp} & \textbf{Overall} \\
\midrule
Naive RAG & 5.1k & 90.0 & \textbf{98.2} & 53.3 & 48.9 & 67.9 & 39.8 & 61.0 \\
Full Context & 105.1k & 87.1 & 89.3 & 36.7 & 45.1 & 76.9 & 31.6 & 56.8 \\
\addlinespace[3pt]
Mem0 & \textbf{1.1k} & 82.9 & 26.8 & 90.0 & 63.2 & 66.7 & 72.2 & 66.4 \\
Zep & 1.6k & 92.9 & 75.0 & 53.3 & 47.4 & 74.4 & 54.1 & 63.2 \\
MemOS & 1.4k & 95.7 & 67.9 & \textbf{96.7} & 70.7 & 74.4 & 77.4 & 77.8 \\
EverMemOS & 2.8k & 97.1 & 85.7 & 93.3 & 73.7 & 89.7 & 77.4 & 83.0 \\
\midrule
\textsc{KITE} & 1.6k & \textbf{98.6} & 85.7 & 63.3 & \textbf{82.0} & \textbf{91.0} & \textbf{85.7} & \textbf{86.0} \\
\bottomrule
\end{tabular}

\vspace{2pt}
\parbox{\linewidth}{
\footnotesize
SS-U, SS-A, and SS-P denote single-session user, assistant, and preference questions; Know and Temp refer to knowledge-update and temporal-reasoning questions.
}
\end{table}

\subsection{Benchmark Performance}
\label{sec:benchmark-performance}

\paragraph{Superior vector-free performance.}
Tables~\ref{tab:locomo} and~\ref{tab:longmemeval} show that \textsc{KITE} achieves the highest overall accuracy among the compared systems without relying on dense embeddings or neural reranking.
It reaches an accuracy of 93.5\% on LoCoMo, outperforming EverMemOS by 0.4 percentage points, and 86.0\% on the LongMemEval-S questions, widening the margin to 3.0 percentage points.
Compared with full-context inference, \textsc{KITE} further improves accuracy by 2.6 and 29.2 percentage points on LoCoMo and LongMemEval-S, respectively.
Overall, these results show that strong long-term memory performance does not require dense similarity retrieval or full-history input to the reader.

\paragraph{Strong temporal and multi-hop reasoning.}
Notably, the strongest advantage of \textsc{KITE} appears in temporal reasoning, where it achieves 92.2\% on LoCoMo and 85.7\% on LongMemEval-S, with the latter exceeding the next-best result by 8.3 percentage points.
This gain is consistent with \textsc{KITE}'s explicit treatment of time, where temporal information is preserved during memory construction and translated into constraints at query time to identify chronologically appropriate evidence.
Beyond temporal reasoning, \textsc{KITE} also excels on multi-hop questions by carrying previously gathered evidence forward across dependent retrieval steps to connect information between different sessions.

\paragraph{Compact reader context.}
Despite its high accuracy, \textsc{KITE} averages 1.5k reader-context tokens per question on LoCoMo and 1.6k on LongMemEval-S.
By comparison, EverMemOS uses 2.3k and 2.8k tokens, respectively, while full-context inference requires 20.3k and 105.1k.
Together, these results show that \textsc{KITE} combines leading overall accuracy with a compact evidence set, substantially reducing the amount of information presented to the reader.

\paragraph{Preference reasoning remains challenging.}
However, \textsc{KITE} is less competitive on single-session preference questions, achieving an accuracy of 63.3\% on the LongMemEval-S benchmark.
This result points to the difficulty of consistently aligning indirect or paraphrased preference expressions with the indexed memory representation.

\subsection{Effectiveness of Deterministic Retrieval}
\label{sec:retrieval-effectiveness}

To isolate the effect of retrieval from memory construction, we freeze the memory store and compare retrieval strategies over the same representations.
We compare dense-vector retrieval, BM25, and their hybrid with \textsc{KITE}'s plan-guided retrieval.
For a controlled comparison, all methods share identical reader settings and reader-context token budgets, with LLM-as-a-judge accuracy and evidence recall measuring answer quality and retrieval effectiveness.

\begin{table}[t]
\centering
\small
\setlength{\tabcolsep}{2.4pt}
\renewcommand{\arraystretch}{1.05}

\newcommand{\headercell}[1]{%
  \begin{tabular}[c]{@{}c@{}}#1\end{tabular}%
}

\begin{minipage}[t]{0.44\linewidth}
\centering
\caption{Accuracy (\%) and recall (\%) of retrieval strategies over fixed memories.}
\label{tab:fixed-memory-retrieval}

\begin{tabular}{@{}lcccc@{}}
\toprule
\multirow{2}{*}[-0.7ex]{%
  \shortstack[l]{\textbf{Retrieval}\\\textbf{Strategy}}}
& \multicolumn{2}{c}{\textbf{LoCoMo}}
& \multicolumn{2}{c@{}}{\textbf{LongMemEval-S}} \\
\cmidrule(lr){2-3}
\cmidrule(lr){4-5}
& \textbf{Acc.}
& \textbf{Rec.}
& \makebox[3.5em][c]{\textbf{Acc.}}
& \makebox[3.5em][c]{\textbf{Rec.}} \\
\midrule
Dense  & 85.4 & 78.1 & 76.9 & 90.5 \\
BM25   & 81.3 & 73.2 & 72.4 & 84.6 \\
Hybrid & 87.8 & 79.9 & 78.0 & 92.8 \\
\midrule
\textsc{KITE}
& \textbf{93.5}
& \textbf{87.4}
& \textbf{86.0}
& \textbf{97.6} \\
\bottomrule
\end{tabular}
\end{minipage}
\hfill%
\begin{minipage}[t]{0.52\linewidth}
\centering
\caption{Comparison of topic vocabulary policies for memory indexing on LoCoMo.}
\label{tab:vocabulary-governance}

\begin{tabular*}{\linewidth}{@{}l@{\extracolsep{\fill}}cccc@{}}
\toprule
\noalign{\vskip 2.4pt}
\headercell{\textbf{Vocabulary Policy}}
& \headercell{\textbf{Avg.}\\\textbf{Topics}}
& \headercell{\textbf{Avg.}\\\textbf{Tokens}}
& \headercell{\textbf{Acc.}\\\textbf{(\%)}}
& \headercell{\textbf{Rec.}\\\textbf{(\%)}} \\
\noalign{\vskip 2.4pt}
\midrule
Fixed Vocabulary
& 12.0 & 2.1k & 91.0 & \textbf{94.2} \\
Immediate Expansion
& 72.4 & \textbf{1.5k} & 87.9 & 78.3 \\
Delayed Expansion
& 62.5 & 1.6k & 91.7 & 82.7 \\
\midrule
\textsc{KITE}
& 58.0 & 1.6k & \textbf{93.5} & 87.4 \\
\bottomrule
\end{tabular*}
\end{minipage}

\end{table}

Table~\ref{tab:fixed-memory-retrieval} shows that \textsc{KITE} achieves the highest accuracy and recall on both benchmarks with memory construction held fixed.
Notably, recall is consistently lower than accuracy on LoCoMo because the gold evidence often includes more dialogue turns than are strictly necessary to answer the question.
A correct answer may therefore be produced without retrieving every annotated turn.
Even so, \textsc{KITE} improves recall over the hybrid baseline by 7.5 percentage points, indicating that \textsc{KITE} retrieves a fuller set of supporting evidence, rather than only the most relevant matches.

By contrast, LongMemEval-S measures session-level recall, which is uniformly higher across methods.
Nevertheless, \textsc{KITE} is consistently stronger in recall and answer accuracy.
This further supports the efficacy of \textsc{KITE}'s deterministic retrieval, which evaluates explicit query conditions before ranking rather than relying on relevance scores alone.

\subsection{Effectiveness of Vocabulary Governance}
\label{sec:vocabulary-governance}

In addition, we examine how governance of the topic vocabulary affects index growth and query compilation on LoCoMo.
All variants start from the same canonical topic roots and process conversations in chronological order.
\emph{Fixed Vocabulary} keeps the initial topic set unchanged, whereas \emph{Immediate Expansion} admits valid new topics as soon as they are proposed.
\emph{Delayed Expansion} keeps proposed topics provisional until they recur across multiple sessions and time periods, while \textsc{KITE} further consolidates redundant sibling topics.
Each variant uses its resulting topic vocabulary for both memory indexing and query compilation, with all other experimental settings remaining fixed.

Table~\ref{tab:vocabulary-governance} illustrates the value of governing topic evolution rather than simply restricting or expanding the vocabulary.
\emph{Immediate Expansion} yields the lowest recall and accuracy, suggesting that related evidence is scattered across too many labels.
Conversely, \emph{Fixed Vocabulary} achieves higher recall, but its longer reader context may include tangential memories that complicate answer generation.
With controlled admission and consolidation, \textsc{KITE} achieves the highest accuracy with fewer topics than expansion-only variants and the same average context length as \emph{Delayed Expansion}.

\begin{table}[t]
\centering
\small
\setlength{\tabcolsep}{4.2pt}
\renewcommand{\arraystretch}{1.05}
\caption{Ablation study of \textsc{KITE} on LoCoMo using LLM-as-a-judge accuracy (\%).}
\label{tab:ablation}

\begin{tabular}{@{}lcccccc@{}}
\toprule
\multirow{2}{*}[-0.7ex]{\textbf{Category}}
& \multicolumn{1}{c}{\textbf{Indexing}}
& \multicolumn{2}{c}{\textbf{Retrieval}}
& \multicolumn{2}{c}{\textbf{Ranking}}
& \multirow{2}{*}[-0.7ex]{\textbf{Full KITE}} \\
\cmidrule(lr){2-2}
\cmidrule(lr){3-4}
\cmidrule(lr){5-6}
& \textbf{w/o Facets}
& \textbf{Indexed Only}
& \textbf{Lexical Only}
& \textbf{Recency}
& \textbf{Neural}
& \\
\midrule
Single   & 93.2 & 93.5 & 85.0 & 64.4 & 94.8 & \textbf{95.2} \\
Multi    & 90.4 & 87.6 & 86.9 & 69.1 & \textbf{95.0} & 93.3 \\
Temporal & 87.9 & 86.3 & 71.3 & 45.5 & 91.6 & \textbf{92.2} \\
Open     & 67.7 & 71.9 & 64.6 & 52.1 & 72.9 & \textbf{83.3} \\
\midrule
\textbf{Overall} & 90.0 & 89.5 & 81.2 & 60.6 & 92.8 & \textbf{93.5} \\
\bottomrule
\end{tabular}
\end{table}

\subsection{Ablation Studies}
\label{sec:ablation}

We conduct ablation studies on LoCoMo to evaluate the main design choices of \textsc{KITE}.
As shown in Table~\ref{tab:ablation}, each experiment changes one part of the pipeline with the rest fixed.
This isolates the contribution of each design choice.

\paragraph{Typed facets.}
Removing typed facets while retaining topic indexing reduces overall accuracy from 93.5\% to 90.0\%.
The largest declines occur on open-domain and temporal questions, at 15.6 and 4.3 percentage points, respectively.
These cases often hinge on distinctions beyond broad thematic alignment, including contextual attributes and temporal cues.
Such facets complement the topic hierarchy with finer-grained constraints for isolating supporting evidence.

\paragraph{Retrieval channels.}
We evaluate both retrieval channels separately and in combination to assess their individual contributions.
Specifically, \emph{Indexed Only} retains only plan-guided retrieval over the governed knowledge index, whereas \emph{Lexical Only} uses the sparse lexical channel alone.
The indexed variant outperforms its lexical counterpart across all question categories, while their combination in full \textsc{KITE} achieves the highest overall accuracy.
These results support the indexed path as the primary retrieval mechanism, with lexical matching providing complementary evidence.

\paragraph{Explicit ranking.}
To assess the role of evidence ranking, we replace the explicit scoring policy with two alternatives.
\emph{Recency} prioritizes memories from newer sessions, whereas \emph{Neural} evaluates query--memory relevance with a pretrained cross-encoder.
The sharp decline under \emph{Recency} indicates that favoring newer memories alone is insufficient for prioritizing the supporting evidence.
Although \emph{Neural} performs better on multi-hop questions, the explicit policy achieves higher overall accuracy by 0.7 percentage points while keeping ranking decisions inspectable.

\section{Conclusion}
\label{sec:conclusion}

We introduced \textsc{KITE}, a vector-free framework that formulates long-term conversational memory retrieval as an explicit decision procedure.
To this end, \textsc{KITE} organizes conversations into structured memories within a governed knowledge index keyed by topic.
At query time, the same index guides the compilation of each question into a retrieval plan, aligning its interpretation with the structure of stored memories.
The plan is then executed deterministically to select evidence that satisfies explicit contextual and temporal conditions.
In this way, retrieval is driven by whether the available information meets the question's requirements, rather than by semantic similarity alone.

Results on LoCoMo and LongMemEval-S show that \textsc{KITE} combines high accuracy with compact reader contexts.
It achieves 93.5\% and 86.0\% using only 1.5k and 1.6k tokens, respectively, with particular strength in temporal and multi-hop reasoning.
With a fixed memory store, plan-guided retrieval excels in answer accuracy and evidence recall, highlighting the role of explicit conditions in recovering supporting information.
Vocabulary-governance experiments further suggest that controlled expansion and consolidation improve accuracy by limiting topic fragmentation.

Further improvements could focus on interpreting indirect preference expressions more consistently across memory construction and query compilation.
These refinements would strengthen the semantic alignment between the two stages while continuing to preserve \textsc{KITE}'s deterministic evidence selection through a shared knowledge index.

\bibliography{references}
\bibliographystyle{kite}

\appendix
\section{LLM Statement}
\label{app:llm-statement}
In this work, we used large language models (LLMs) for three purposes.
First, \texttt{gpt-4.1-mini} was integrated into our experimental framework as an LLM component during system execution.
Second, LLMs were used sparingly to identify and retrieve relevant literature; all cited literature was independently reviewed and verified by the authors.
Finally, we used LLMs for language polishing, limited to correcting grammatical errors and improving the clarity and fluency of the text.
The research questions, methodology, experimental design, analysis, and conclusions were developed by the authors.
All AI-assisted work was reviewed by the authors, who take full responsibility for the final content of this work.
No confidential or personal data were provided to external LLM services at any point.

\section{Experimental Details}
\label{app:experimental-details}

\subsection{Benchmarks and Evaluation Metrics}
\label{app:evaluation-metrics}

\paragraph{Benchmarks.}
Our LoCoMo evaluation \citep{maharana2024evaluating} covers 1,540 questions across ten multi-session conversations, including single-hop (841), multi-hop (282), temporal (321), and open-domain (96) reasoning.
The LongMemEval-S benchmark \citep{wu2024longmemeval} comprises a total of 500 questions posed over substantially longer interaction histories, spanning single-session user information (70), single-session assistant information (56), single-session preferences (30), multi-session reasoning (133), knowledge updates (78), and temporal reasoning (133).

\paragraph{Reader-context tokens.}
Avg.\ Tokens denotes the average number of context tokens provided to the answer reader per question, counted using the \texttt{o200k\_base} tokenizer.
Lower values indicate a more compact context.
Notably, this metric measures only the reader-context size rather than the total number of tokens consumed by the system.

\paragraph{Answer accuracy.}
LLM-as-a-judge accuracy is the percentage of generated answers judged correct by an LLM evaluator.
For \textsc{KITE}, we use \texttt{gpt-4.1-mini} as the judge and report accuracy overall and by question category.

\paragraph{Evidence recall.}
Evidence recall measures the fraction of annotated supporting evidence recovered by retrieval.
For a question $q$, recall is computed as
\begin{equation}
    r_q = \frac{|G_q \cap R_q|}{|G_q|},
    \qquad |G_q| > 0,
    \label{eq:appendix-recall}
\end{equation}
where $G_q$ and $R_q$ denote the sets of annotated and retrieved evidence units, respectively.
Units are LoCoMo dialogue turns and LongMemEval-S sessions.
We report the average $r_q$ over questions with annotated supporting evidence.

\subsection{Baseline Evaluation Settings}
\label{app:baseline-settings}

\paragraph{Memory-system comparison.}
For consistency with recent published evaluations, we base our comparison with Mem0, Zep, MemOS, and EverMemOS on the results consolidated by EverMemOS \citep{hu2026evermemos}.
We follow EverMemOS in using \texttt{gpt-4.1-mini} for final answer generation, in keeping with the reference model setting.

\paragraph{Full context.}
Full Context bypasses memory construction and retrieval by providing the complete conversation history directly to the answer reader.
Sessions are presented chronologically, with session dates and speaker labels preserved.
No retrieval-specific filtering, summarization, or reranking is applied to the full conversation history.

\paragraph{Naive RAG.}
We implement Naive RAG using the standard vector-store pipeline provided
by LlamaIndex\footnote{\url{https://github.com/run-llama/llama_index}}.
Each dated conversation session is treated as a retrieval unit and is embedded independently using \texttt{text-embedding-3-small}.
At query time, the question is embedded with the same model, and the highest-scoring sessions by cosine similarity are supplied to the answer reader in chronological order, with session dates and speaker labels preserved.
No memory extraction, query expansion, temporal filtering, or neural reranking is applied.

\subsection{\textsc{KITE} Implementation Settings}
\label{app:kite-settings}

\paragraph{Models and decoding.}
We use \texttt{gpt-4.1-mini} throughout all model-based stages of \textsc{KITE}, including memory extraction, vocabulary consolidation, query compilation, and final answer generation.
All model calls use a temperature of zero, except query compilation, which draws one plan at temperature zero and two further candidates at temperature 0.8. 
The same model and decoding settings are employed for both LoCoMo and LongMemEval-S.

\paragraph{Initial vocabularies and governed expansion.}
\textsc{KITE} starts from a small set of broad topic roots for each benchmark.
Table~\ref{tab:kite-initial-vocab} lists these roots, the available memory-entry types, and the event facets shared by both benchmarks.
Entities, participants, objects, and places are extracted from the conversations themselves rather than predefined.

\begin{table}[h]
\centering
\small
\setlength{\tabcolsep}{4pt}
\renewcommand{\arraystretch}{1.05}
\caption{Complete initial vocabularies used to construct the knowledge index.}
\label{tab:kite-initial-vocab}
\begin{tabularx}{\linewidth}{@{}l>{\raggedright\arraybackslash}X@{}}
\toprule
\textbf{Vocabulary} & \textbf{Values} \\
\midrule
LoCoMo topic roots &
relationships, family, work and career, education, health, activities and hobbies, arts and creativity, travel, life events, identity and beliefs, possessions, and community \\
\addlinespace[2pt]
LoCoMo entry types &
event, plan, preference, identity, relationship, emotion, opinion, health, possession, and other \\
\addlinespace[2pt]
LongMemEval-S topic roots &
personal life, work and career, health and fitness, travel, shopping and finance, food and cooking, family and relationships, hobbies and entertainment, technical help, writing tasks, education and learning, and home and lifestyle \\
\addlinespace[2pt]
LongMemEval-S entry types &
event, plan, preference, identity, relationship, opinion, task request, knowledge, and other \\
\addlinespace[2pt]
Shared event facets &
trip, move, purchase, sale, injury, illness, recovery, award, competition, performance, graduation, job change, wedding, birth, death, adoption, meeting, celebration, accident, release, rejection, milestone, and first-time event \\
\bottomrule
\end{tabularx}
\end{table}

During memory construction, the extractor reuses the most specific suitable topic and creates a new child topic only when needed.
A new topic remains provisional until it appears in three sessions across at least two different months, while redundant sibling topics are merged only when repeated consolidation reviews agree.
LongMemEval-S additionally applies a final reassignment pass that moves entries from broad roots to suitable established child topics.

\paragraph{Evidence ranking and channel fusion.}
A validated retrieval plan first filters the memory entries according to its explicit conditions.
The eligible entries are then ordered by a fixed additive score based on their topic, entity, and facet matches, with the contributions summarized in Table~\ref{tab:kite-ranking}.
The indexed ranking is complemented by a sparse lexical ranking that favors rare matching terms and discounts incidental matches in longer entries.
Questions containing an explicit date or month also receive a date-based ranking over entries from the corresponding period.
The available rankings are combined using reciprocal rank fusion,
\begin{equation}
F(m)
=
\sum_{c\in\mathcal{R}(m)}
\frac{1}{60+r_c(m)},
\label{eq:retrieval-fusion}
\end{equation}
where $\mathcal{R}(m)$ denotes the set of channels that retrieve $m$ and $r_c(m)$ is the zero-based rank of $m$ in channel $c$.
The fusion policy remains fixed across both benchmarks and uses neither dense embeddings nor neural reranking.

\begin{table}[t]
\centering
\small
\setlength{\tabcolsep}{5pt}
\renewcommand{\arraystretch}{1.05}
\caption{Fixed score contributions used to rank entries selected by a structured retrieval plan.}
\label{tab:kite-ranking}
\begin{tabular}{@{}lc@{}}
\toprule
\textbf{Matching signal} & \textbf{Score contribution} \\
\midrule
Exact topic match & $3.0$ \\
More specific descendant topic & $2.0$ \\
Broad hierarchical topic match & $0.8$ \\
Exact entity match & $2.0$ \\
Matched facet type & $1.5$ \\
Memory-entry type match & $1.0$ \\
Participant match & $0.5$ \\
Confidence adjustment & $0.0$--$0.4$ \\
Rare entity of the requested type & up to $3.5$ \\
Lexical condition & up to $6.0$ \\
\bottomrule
\end{tabular}
\end{table}

\paragraph{Evidence packing and answer generation.}
After channel fusion, duplicate entries are removed and the remaining items are added in fused rank order until a fixed token budget is reached.
The selected material is then arranged in chronological order and retains dates, speaker labels, short source quotations, and provenance information when available.
Given the same index and validated plan, this process produces the same supporting material every time.
The answer reader sees only the question and the assembled context, without access to the complete conversation history.
A bounded follow-up check then handles responses that are incomplete or weakly supported.
If the first reading identifies a specific missing fact, the system may run one additional search over the same index.
No external information is introduced, and the final response follows each benchmark's rules for answering or abstaining.

\section{Prompt Templates}
\label{app:prompts}

The templates below illustrate the main instructions and output formats used by \textsc{KITE}.\footnote{The full prompts used in our experiments are provided in the released code.}
Braces mark runtime inputs.

\subsection{Memory Extraction Prompt}
\label{app:prompt-memory}

The extraction prompt converts each dated session into source-linked facts with topic assignments and temporal information.
LoCoMo emphasizes coverage of the dialogue, while LongMemEval-S prioritizes the user's statements.

\begin{kiteprompt}{Prompt 1: Memory extraction}
Encode this conversation session into self-contained, atomic facts.
SESSION DATE: {date} ({weekday}), {time}
SPEAKERS: {speakers}
TOPIC TAXONOMY (codes ending in ? are provisional): {tree}
KNOWN ENTITIES: {entities}
DIALOG (id|speaker||text):
{dialog}

Rules:
- Cover substantive details without inferring unstated facts.
  Preserve names, counts, amounts, and other askable details.
- Write one self-contained claim per fact. Resolve month and year
  references against the session date, but leave weekday references
  relative. Record the event time in "t" when known.
- Set "kind" to one of {kinds}. Use "who" for the person or role
  the fact is about, which may differ from the speaker.
  Set "conf" to high, med, or low.
- Assign one to three of the most specific applicable topic codes.
  Reuse existing codes when possible. If only a broad root applies,
  propose a specific child under that root.
- Record named entities and the dialogue IDs that directly support
  each fact in "entities" and "src".

Return JSON only:
{
  "facts": [{
    "content": str, "kind": str,
    "who": str, "t": str | null, "conf": "high|med|low",
    "topics": [str], "entities": [str], "src": [str]
  }],
  "proposals": [{"code": str, "parent": str}],
  "entity_types": {"name": "type"}
}
\end{kiteprompt}

\subsection{Topic Vocabulary Consolidation Prompt}
\label{app:prompt-topic-consolidation}

The consolidation prompt reviews potentially redundant sibling topics in the governed topic vocabulary.
It identifies equivalent topics and selects a canonical topic for merging, while preserving distinctions useful for retrieval.

\begin{kiteprompt}{Prompt 2: Topic vocabulary consolidation}
You maintain the controlled topic taxonomy for a long-term
conversational memory. For each pair of topic codes, decide
whether they denote exactly the same conversational subject.

Merge only true synonyms, spelling variants, or equivalent
phrasings. Do not merge a broad topic with one of its subtopics,
a whole with one of its parts, or two subjects that are merely
related. Preserve distinctions that would help a later retrieval
request identify the right conversation. When merging, choose
the clearer, more specific code that remains understandable on
its own.

TOPIC PAIRS:
{pairs}

Return JSON only:
{"decisions": [{"i": 0, "merge": true, "winner": str}]}
\end{kiteprompt}

\subsection{Index-Grounded Query Compilation Prompt}
\label{app:prompt-query-compilation}

The compilation prompt maps a natural-language question to a plan using the current topic vocabulary and entity index.
They allow independent subqueries or dependent stages in which an earlier result narrows a later search.

\begin{kiteprompt}{Prompt 3: Index-grounded query compilation}
Translate the question into a retrieval plan (JSON).
TOPIC TAXONOMY: {tree}
ENTITIES: {entities}
FACT KINDS: {kinds}
UNITS (id=date): {units}
TODAY: {today}
QUESTION: {question}

Use one to three simple subqueries and prefer existing topic
and entity codes. Include a broad fact grep query using
distinctive word stems. Use named stages when a later search
depends on an earlier result.

Parallel plan:
{
  "queries": [{
    "select": "facts|lines|units",
    "where": {
      "grep": str, "who": [str],
      "topics": [{"code": str, "closure": true}],
      "entities": [str], "kind": [str], "time": [str, str]
    },
    "pipe": [{"op": "head", "n": 5}]
  }],
  "intent": "factual"
}

Staged plan:
{
  "stages": [
    {"name": "anchor", "select": "units",
     "where": {"grep": str},
     "pipe": [{"op": "head", "n": 1}]},
    {"select": "facts",
     "where": {"units": "$anchor.units"}}
  ],
  "intent": "factual"
}
Return JSON only.
\end{kiteprompt}
\end{document}

%% file: math_commands.tex
\usepackage{amsmath,amsfonts,bm}

\def\eqref#1{equation~\ref{#1}}

\def\1{\bm{1}}

\DeclareMathAlphabet{\mathsfit}{\encodingdefault}{\sfdefault}{m}{sl}
\SetMathAlphabet{\mathsfit}{bold}{\encodingdefault}{\sfdefault}{bx}{n}













%% file: kite.bbl
\begin{thebibliography}{32}
\providecommand{\natexlab}[1]{#1}
\providecommand{\url}[1]{\texttt{#1}}
\expandafter\ifx\csname urlstyle\endcsname\relax
  \providecommand{\doi}[1]{doi: #1}\else
  \providecommand{\doi}{doi: \begingroup \urlstyle{rm}\Url}\fi

\bibitem[Abtahi et~al.(2026)Abtahi, Rahnema, Patel, Patel, Fekri, and
  Khani]{abtahi2026memanto}
Seyed~Moein Abtahi, Rasa Rahnema, Hetkumar Patel, Neel Patel, Majid Fekri, and
  Tara Khani.
\newblock Memanto: Typed semantic memory with information-theoretic retrieval
  for long-horizon agents.
\newblock \emph{arXiv preprint arXiv:2604.22085}, 2026.

\bibitem[Beltagy et~al.(2020)Beltagy, Peters, and Cohan]{beltagy2020longformer}
Iz~Beltagy, Matthew~E. Peters, and Arman Cohan.
\newblock Longformer: The long-document transformer.
\newblock \emph{arXiv preprint arXiv:2004.05150}, 2020.

\bibitem[Bulatov et~al.(2022)Bulatov, Kuratov, and
  Burtsev]{bulatov2022recurrent}
Aydar Bulatov, Yury Kuratov, and Mikhail Burtsev.
\newblock Recurrent memory transformer.
\newblock \emph{Advances in Neural Information Processing Systems},
  35:\penalty0 11079--11091, 2022.

\bibitem[Chhikara et~al.(2025)Chhikara, Khant, Aryan, Singh, and
  Yadav]{chhikara2025mem0}
Prateek Chhikara, Dev Khant, Saket Aryan, Taranjeet Singh, and Deshraj Yadav.
\newblock {Mem0}: Building production-ready {AI} agents with scalable long-term
  memory.
\newblock In \emph{ECAI 2025: 28th European Conference on Artificial
  Intelligence}, volume 413 of \emph{Frontiers in Artificial Intelligence and
  Applications}, pp.\  2993--3000. IOS Press, 2025.

\bibitem[Ge et~al.(2025)Ge, Romeo, Cai, Shu, Benajiba, Sunkara, and
  Zhang]{ge2025tremu}
Yubin Ge, Salvatore Romeo, Jason Cai, Raphael Shu, Yassine Benajiba, Monica
  Sunkara, and Yi~Zhang.
\newblock {TReMu}: Towards neuro-symbolic temporal reasoning for {LLM}-agents
  with memory in multi-session dialogues.
\newblock In \emph{Findings of the Association for Computational Linguistics:
  ACL 2025}, pp.\  18974--18988, 2025.

\bibitem[Goswami et~al.(2025)Goswami, Wang, Twardowski, and van~de
  Weijer]{goswami2025query}
Dipam Goswami, Liying Wang, Bart{\l}omiej Twardowski, and Joost van~de Weijer.
\newblock Query drift compensation: Enabling compatibility in continual
  learning of retrieval embedding models.
\newblock In \emph{Proceedings of The 4th Conference on Lifelong Learning
  Agents}, volume 330 of \emph{Proceedings of Machine Learning Research}, pp.\
  324--340. PMLR, 2025.

\bibitem[Hsieh et~al.(2024)Hsieh, Sun, Kriman, Acharya, Rekesh, Jia, Zhang, and
  Ginsburg]{hsieh2024ruler}
Cheng-Ping Hsieh, Simeng Sun, Samuel Kriman, Shantanu Acharya, Dima Rekesh, Fei
  Jia, Yang Zhang, and Boris Ginsburg.
\newblock {RULER}: What's the real context size of your long-context language
  models?
\newblock In \emph{First Conference on Language Modeling}, 2024.

\bibitem[Hu et~al.(2023)Hu, Fu, Du, Luo, Zhao, and Zhao]{hu2023chatdb}
Chenxu Hu, Jie Fu, Chenzhuang Du, Simian Luo, Junbo Zhao, and Hang Zhao.
\newblock {ChatDB}: Augmenting {LLMs} with databases as their symbolic memory.
\newblock \emph{arXiv preprint arXiv:2306.03901}, 2023.

\bibitem[Hu et~al.(2026{\natexlab{a}})Hu, Gao, Zhou, Xu, Bai, Li, Zhang, Li,
  Zhang, Bing, and Deng]{hu2026evermemos}
Chuanrui Hu, Xingze Gao, Zuyi Zhou, Dannong Xu, Yi~Bai, Xintong Li, Hui Zhang,
  Tong Li, Chong Zhang, Lidong Bing, and Yafeng Deng.
\newblock {EverMemOS}: A self-organizing memory operating system for structured
  long-horizon reasoning.
\newblock \emph{arXiv preprint arXiv:2601.02163}, 2026{\natexlab{a}}.

\bibitem[Hu et~al.(2026{\natexlab{b}})Hu, Teng, Zhang, He, and
  Fan]{hu2026querylink}
Xuxian Hu, Zhu Teng, Wei Zhang, Ming He, and Jianping Fan.
\newblock {QueryLink}: Leveraging query-memory alignment for long-term
  reasoning in {LLM} agents.
\newblock In \emph{Findings of the Association for Computational Linguistics:
  ACL 2026}, pp.\  15608--15621, 2026{\natexlab{b}}.

\bibitem[Hutter et~al.(2025)Hutter, Rau, Marx, and Kamps]{hutter2025lost}
Jan Hutter, David Rau, Maarten Marx, and Jaap Kamps.
\newblock Lost but not only in the middle: Positional bias in retrieval
  augmented generation.
\newblock In \emph{European Conference on Information Retrieval}, pp.\
  247--261. Springer, 2025.

\bibitem[Kang et~al.(2025)Kang, Ji, Zhao, and Bai]{kang2025memory}
Jiazheng Kang, Mingming Ji, Zhe Zhao, and Ting Bai.
\newblock Memory {OS} of {AI} agent.
\newblock In \emph{Proceedings of the 2025 Conference on Empirical Methods in
  Natural Language Processing}, pp.\  25961--25970, 2025.

\bibitem[Kim et~al.(2025)Kim, Kim, Lee, and Kim]{kim2025speak}
Hyuhng~Joon Kim, Youna Kim, Sang-goo Lee, and Taeuk Kim.
\newblock When to speak, when to abstain: Contrastive decoding with abstention.
\newblock In \emph{Proceedings of the 63rd Annual Meeting of the Association
  for Computational Linguistics (Volume 1: Long Papers)}, pp.\  9710--9730,
  2025.

\bibitem[Li et~al.(2025)Li, Xi, Li, Chen, Chen, Song, Niu, Wang, Yang, Tang,
  et~al.]{li2025memos}
Zhiyu Li, Chenyang Xi, Chunyu Li, Ding Chen, Boyu Chen, Shichao Song, Simin
  Niu, Hanyu Wang, Jiawei Yang, Chen Tang, et~al.
\newblock {MemOS}: A memory {OS} for {AI} system.
\newblock \emph{arXiv preprint arXiv:2507.03724}, 2025.

\bibitem[Liang et~al.(2025)Liang, Bo, Gui, Zhu, Zhong, Zhao, Sun, Zhang, Zhou,
  Chen, Zhang, and Chen]{liang2025kag}
Lei Liang, Zhongpu Bo, Zhengke Gui, Zhongshu Zhu, Ling Zhong, Peilong Zhao,
  Mengshu Sun, Zhiqiang Zhang, Jun Zhou, Wenguang Chen, Wen Zhang, and Huajun
  Chen.
\newblock {KAG}: Boosting {LLMs} in professional domains via knowledge
  augmented generation.
\newblock In \emph{Companion Proceedings of the ACM on Web Conference 2025},
  pp.\  334--343, 2025.

\bibitem[Liu et~al.(2024)Liu, Lin, Hewitt, Paranjape, Bevilacqua, Petroni, and
  Liang]{liu2024lost}
Nelson~F. Liu, Kevin Lin, John Hewitt, Ashwin Paranjape, Michele Bevilacqua,
  Fabio Petroni, and Percy Liang.
\newblock Lost in the middle: How language models use long contexts.
\newblock \emph{Transactions of the Association for Computational Linguistics},
  12:\penalty0 157--173, 2024.

\bibitem[Long et~al.(2026)Long, Chen, Zeng, Wang, Guo, and
  Tang]{long2026memtrace}
Xianxuan Long, Zhikai Chen, Shenglai Zeng, Shouren Wang, Kai Guo, and Jiliang
  Tang.
\newblock {MemTrace}: Probing what final accuracy misses in long-term memory.
\newblock \emph{arXiv preprint arXiv:2606.17328}, 2026.

\bibitem[Maharana et~al.(2024)Maharana, Lee, Tulyakov, Bansal, Barbieri, and
  Fang]{maharana2024evaluating}
Adyasha Maharana, Dong-Ho Lee, Sergey Tulyakov, Mohit Bansal, Francesco
  Barbieri, and Yuwei Fang.
\newblock Evaluating very long-term conversational memory of {LLM} agents.
\newblock In \emph{Proceedings of the 62nd Annual Meeting of the Association
  for Computational Linguistics (Volume 1: Long Papers)}, pp.\  13851--13870,
  2024.

\bibitem[Mo et~al.(2025)Mo, Yu, Kazdan, Cabezas, Mpala, Yu, Cundy, Kanatsoulis,
  and Koyejo]{mo2026kggen}
Belinda Mo, Kyssen Yu, Joshua Kazdan, Joan Cabezas, Proud Mpala, Lisa Yu, Chris
  Cundy, Charilaos Kanatsoulis, and Sanmi Koyejo.
\newblock {KGGen}: Extracting knowledge graphs from plain text with language
  models.
\newblock \emph{Advances in Neural Information Processing Systems},
  38:\penalty0 30092--30115, 2025.

\bibitem[Monath et~al.(2024)Monath, Grathwohl, Boratko, Fergus, McCallum, and
  Zaheer]{monath2024fresh}
Nicholas Monath, Will~Sussman Grathwohl, Michael Boratko, Rob Fergus, Andrew
  McCallum, and Manzil Zaheer.
\newblock A fresh take on stale embeddings: Improving dense retriever training
  with corrector networks.
\newblock In \emph{Proceedings of the 41st International Conference on Machine
  Learning}, volume 235 of \emph{Proceedings of Machine Learning Research},
  pp.\  36072--36087. PMLR, 2024.

\bibitem[Packer et~al.(2023)Packer, Wooders, Lin, Fang, Patil, Stoica, and
  Gonzalez]{packer2023memgpt}
Charles Packer, Sarah Wooders, Kevin Lin, Vivian Fang, Shishir~G. Patil, Ion
  Stoica, and Joseph~E. Gonzalez.
\newblock {MemGPT}: Towards {LLMs} as operating systems.
\newblock \emph{arXiv preprint arXiv:2310.08560}, 2023.

\bibitem[Rasmussen et~al.(2025)Rasmussen, Paliychuk, Beauvais, Ryan, and
  Chalef]{rasmussen2025zep}
Preston Rasmussen, Pavlo Paliychuk, Travis Beauvais, Jack Ryan, and Daniel
  Chalef.
\newblock Zep: a temporal knowledge graph architecture for agent memory.
\newblock \emph{arXiv preprint arXiv:2501.13956}, 2025.

\bibitem[Robertson \& Zaragoza(2009)Robertson and Zaragoza]{robertson2009bm25}
Stephen Robertson and Hugo Zaragoza.
\newblock The probabilistic relevance framework: {BM25} and beyond.
\newblock \emph{Foundations and Trends in Information Retrieval}, 3\penalty0
  (4):\penalty0 333--389, 2009.

\bibitem[Ru et~al.(2024)Ru, Qiu, Hu, Zhang, Shi, Chang, Jiayang, Wang, Sun, Li,
  Zhang, Wang, Jiang, He, Wang, Liu, Zhang, and Zhang]{ru2024ragchecker}
Dongyu Ru, Lin Qiu, Xiangkun Hu, Tianhang Zhang, Peng Shi, Shuaichen Chang,
  Cheng Jiayang, Cunxiang Wang, Shichao Sun, Huanyu Li, Zizhao Zhang, Binjie
  Wang, Jiarong Jiang, Tong He, Zhiguo Wang, Pengfei Liu, Yue Zhang, and Zheng
  Zhang.
\newblock {RAGChecker}: A fine-grained framework for diagnosing
  retrieval-augmented generation.
\newblock \emph{Advances in Neural Information Processing Systems},
  37:\penalty0 21999--22027, 2024.

\bibitem[Trappolini et~al.(2026)Trappolini, Cuconasu, Filice, Maarek, and
  Silvestri]{trappolini2026redefining}
Giovanni Trappolini, Florin Cuconasu, Simone Filice, Yoelle Maarek, and
  Fabrizio Silvestri.
\newblock Redefining retrieval evaluation in the era of {LLMs}.
\newblock In \emph{Proceedings of the 19th Conference of the European Chapter
  of the Association for Computational Linguistics (Volume 1: Long Papers)},
  pp.\  8359--8375, 2026.

\bibitem[Wang et~al.(2026)Wang, Yang, Chen, Xu, Zou, Tang, Xiong, Chen, and
  Li]{wang2026text2mem}
Leo Wang, Lihai Yang, Boyu Chen, Kerun Xu, Gongyi Zou, Bo~Tang, Feiyu Xiong,
  Siheng Chen, and Zhiyu Li.
\newblock {Text2Mem}: A unified memory operation language for memory operating
  system.
\newblock In \emph{Findings of the Association for Computational Linguistics:
  ACL 2026}, pp.\  2105--2119, 2026.

\bibitem[Weller et~al.(2026)Weller, Boratko, Naim, and
  Lee]{weller2026theoretical}
Orion Weller, Michael Boratko, Iftekhar Naim, and Jinhyuk Lee.
\newblock On the theoretical limitations of embedding-based retrieval.
\newblock In \emph{International Conference on Learning Representations}, pp.\
  11202--11216, 2026.

\bibitem[Wu et~al.(2025)Wu, Wang, Yu, Zhang, Chang, and Yu]{wu2024longmemeval}
Di~Wu, Hongwei Wang, Wenhao Yu, Yuwei Zhang, Kai-Wei Chang, and Dong Yu.
\newblock {LongMemEval}: Benchmarking chat assistants on long-term interactive
  memory.
\newblock In \emph{International Conference on Learning Representations}, pp.\
  86809--86836, 2025.

\bibitem[Xu et~al.(2025)Xu, Liang, Mei, Gao, Tan, and Zhang]{xu2026mem}
Wujiang Xu, Zujie Liang, Kai Mei, Hang Gao, Juntao Tan, and Yongfeng Zhang.
\newblock {A-Mem}: Agentic memory for {LLM} agents.
\newblock \emph{Advances in Neural Information Processing Systems},
  38:\penalty0 17577--17604, 2025.

\bibitem[Zhang \& Soh(2024)Zhang and Soh]{zhang2024extract}
Bowen Zhang and Harold Soh.
\newblock Extract, define, canonicalize: An {LLM}-based framework for knowledge
  graph construction.
\newblock In \emph{Proceedings of the 2024 Conference on Empirical Methods in
  Natural Language Processing}, pp.\  9820--9836, 2024.

\bibitem[Zhong et~al.(2024)Zhong, Guo, Gao, Ye, and Wang]{zhong2024memorybank}
Wanjun Zhong, Lianghong Guo, Qiqi Gao, He~Ye, and Yanlin Wang.
\newblock {MemoryBank}: Enhancing large language models with long-term memory.
\newblock \emph{Proceedings of the AAAI Conference on Artificial Intelligence},
  38\penalty0 (17):\penalty0 19724--19731, 2024.

\bibitem[Zhou et~al.(2026)Zhou, Qu, Wu, Kim, Prakash, Rus, Low, and
  Liang]{zhou2026mem1}
Zijian Zhou, Ao~Qu, Zhaoxuan Wu, Sunghwan Kim, Alok Prakash, Daniela Rus, Bryan
  Kian~Hsiang Low, and Paul Liang.
\newblock {MEM1}: Learning to synergize memory and reasoning for efficient
  long-horizon agents.
\newblock In \emph{International Conference on Learning Representations}, pp.\
  58413--58438, 2026.

\end{thebibliography}
